\documentclass[prd,twocolumn]{revtex4}
\usepackage{amsmath}
\usepackage{epsfig}
\begin{document}
\def\b{\bar}
\def\d{\partial}
\def\D{\Delta}
\def\cD{{\cal D}}
\def\cK{{\cal K}}
\def\f{\varphi}
\def\g{\gamma}
\def\G{\Gamma}
\def\l{\lambda}
\def\L{\Lambda}
\def\M{{\Cal M}}
\def\m{\mu}
\def\n{\nu}
\def\p{\psi}
\def\q{\b q}
\def\r{\rho}
\def\t{\tau}
\def\x{\phi}
\def\X{\~\xi}
\def\~{\widetilde}
\def\h{\eta}
\def\bZ{\bar Z}
\def\cY{\bar Y}
\def\bY3{\bar Y_{,3}}
\def\Y3{Y_{,3}}
\def\z{\zeta}
\def\Z{{\b\zeta}}
\def\Y{{\bar Y}}
\def\cZ{{\bar Z}}
\def\`{\dot}
\def\be{\begin{equation}}
\def\ee{\end{equation}}
\def\bea{\begin{eqnarray}}
\def\eea{\end{eqnarray}}
\def\half{\frac{1}{2}}
\def\fn{\footnote}
\def\bh{black hole \ }
\def\cL{{\cal L}}
\def\cH{{\cal H}}
\def\cF{{\cal F}}
\def\cP{{\cal P}}
\def\cM{{\cal M}}
\def\ik{ik}
\def\mn{{\mu\nu}}
\def\a{\alpha}

\title{Beam-like Excitations of Kerr-Schild Geometry and Semiclassical
Mechanism  of Black-Hole Evaporation.}

\author{Alexander Burinskii }

\affiliation{Gravity Research Group, NSI, Russian Academy of
Sciences, B. Tulskaya 52  Moscow 115191 Russia}

\begin{abstract}
It has been observed that exact solutions for electromagnetic (EM)
excitations of the Kerr-Schild (KS) geometry form outgoing beams
which have very strong back reaction to metric and break the black
hole  horizon. As a result, interaction of a black hole with
nearby electromagnetic field and electromagnetic vacuum has to
cover the horizon by a set of fluctuating microholes. We integrate
and analyze the Debney-Kerr-Schild equations for electromagnetic
excitations of a black-hole and obtain that the exact solutions for
outgoing radiation contain two related but radically different
components which shed light on a possible semi-classical
mechanism of black-hole evaporation: a) first component consists
of the singular beam pulses which perforate horizon, breaking its
impenetrability, and  b) another component is regular and
responsible for the loss of mass similar to the known
Vaidya `shining star' radiation. We show also that the mysterious
twosheeted twistor structure of the Kerr-Schild geometry corresponds
to a holographic structure of quantum black hole spacetimes predicted
by Stephens, t' Hooft and Whiting. The resulting Kerr-Schild geometry
of fluctuating twistor-beams takes an intermediate position between the
classical and quantum gravity.

\end{abstract}


\maketitle

\section{Introduction} Diversity of the
recent ideas on the origin of black hole  evaporation \cite{Carl0} has a common
core based on  complex analyticity and conformal field theory, which unifies
the black hole physics with (super)string theory and physics of elementary particles,
as it was first noticed by `t Hooft in \cite{Hooft1}.
These relationships open a way to quantum gravity by a holographic correspondence
between a classical bulk gravity and a quantum Conformal Field Theory which lives
on a holographically dual to bulk 2D boundary \cite{Carl,Stro,Solod,Star}.

In this paper we show that analytic twosheeted structure of the
Kerr-Schild geometry is in perfect agreement with the required
holographic structure of quantum black hole spacetimes
\cite{Hoof,StHW}. Obtaining exact solutions of Debney-Kerr-Schild
equations \cite{DKS} for electromagnetic  excitations of the
Kerr-Schild geometry and their back reaction to metric and
horizon, we arrive at a semiclassical fluctuating Kerr-Schild
geometry which takes an intermediate position between the
classical and quantum gravity. This geometry has a classical
fluctuating fine-grained structure which shows that the usual
representations on the structure of BH and the horizon may be
naive and very far from reality.\fn{Similar consequences are going
from the site of string theory \cite{Mathur}.}

 The usual statements
on stability of the black hole horizon are based on the theorems
(Robinson and Carter)\cite{Chandra} claiming the uniqueness of the
Kerr solution. These theorems are valid under a series of
conditions: stationarity, axial symmetry and asymptotic flatness
of the metric which turn out to be broken under electromagnetic
(EM) excitations of black hole \cite{BurA}. Another vulnerable
point is the traditional use of perturbative approach. The recent
analysis of the exact nonstationary electromagnetic solutions,
performed in the Kerr-Schild formalism \cite{EM3p,BEHM3}, showed
that excitations of black hole do not contain the smooth spherical
harmonics used by the perturbative analysis: elementary exact
electromagnetic excitations on the Kerr background have the form
of singular beams which may be outgoing in any angular direction
which have very strong back reaction on metric and the horizon.
The beams break horizon topologically \cite{BEHM2,BEHM3}, forming
the holes which allow matter to escape interior of black hole.
Origin of this effect lies in analyticity of the Kerr-Schild
solutions and, in particular, in twistor analyticity of the Kerr
congruence determined by the Kerr theorem
\cite{KraSte,BurTwi,Multiks}. The Kerr-Schild form of metric \be
g_\mn =\eta_\mn + 2H k_\m k_\n, \label{KS}\ee has many advantages
with respect to other representations. First of them is the rigid
connection of coordinates $x^\m$ to auxiliary Minkowski space-time
with metric $\eta_\mn $ and the corresponding unfastening of the
coordinates and solutions from positions of the horizons, which
allows one to analyze deformations of the horizon. Second, the
resulting Kerr-Schild solutions are not singular at the horizon.
Third advantage is related with the Kerr-Schild twistor structure
\cite{BurTwi} and correspondence with the requirements of
holographic principle and the presumable properties of a quantum
black hole space-time described by Stephens, t' Hooft and Whiting
in \cite{StHW}.  The Kerr-Schild holographic projection is
realized by the Kerr congruence and generalizes the holographic
map considered by Bousso \cite{Bousso}. The twosheeted Kerr-Schild
structure perfectly matches with holographic approach and allows
one to consider evaporation as a scattering of in-vacuum on a
holographically dual 2+1 dimensional Kerr source which separates
the in and out regions \cite{StHW}, and we obtain that the exact
outgoing solutions have two related but radically different
components which shed light on a possible semi-classical mechanism
of black-hole evaporation:

a) a casual set of singular beam pulses which perforate horizon, breaking its
impenetrability, and

b) a regular component which is responsible for the loss of mass and akin
to the known Vaidya `shining star' radiation.

For the reader convenience we
give here some digression with description of the Kerr-Schild relations \cite{DKS}.
Analyticity of the Kerr-Schild
geometry originates from the complex function $Y = e^{i\phi} \tan
\frac \theta 2$ which is a projection of celestial sphere $S^2$ on
the complex plane. This function determines the Kerr congruence
and complex tetrad forms. The Kerr theorem sets the
dependence $Y(x),$ and the null vector field $k^\m$
of the Kerr-Schild metric form is expressed via function $Y(x)$ \be k_\m dx^\m =
P^{-1}
 (du + \bar Y d \zeta + Y d \bar\zeta - Y \bar Y dv ),
\label{e3}\ee in the null Cartesian coordinates $ 2^{\frac 12}\z =
x+iy , \quad 2^{\frac 1 2} \Z = x-iy , \quad 2^{\frac 12}u = z - t
,\quad 2^{\frac 1 2}v = z + t . $ Therefore, the field
  $k^\m(x), \ x\in M^4$ determines  symmetry of space-time,
polarization of the Kerr-Newman electromagnetic field, direction of
gravitational `dragging` and so on. This vector field is tangent to the Kerr
congruence which is the family of the light-like geodesic lines,
in fact twistors. Twisting structure of the Kerr congruence is shown in Fig.1

\begin{figure}[ht]
\centerline{\epsfig{figure=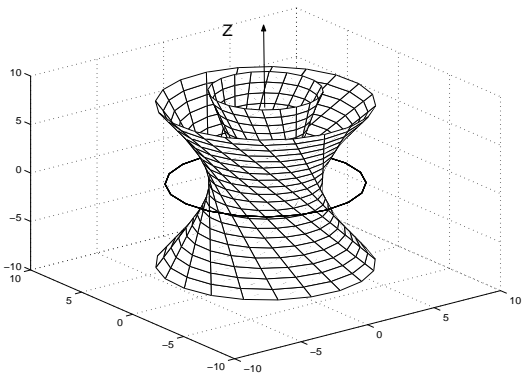,height=5cm,width=7cm}}
\caption{The Kerr singular ring and the Kerr congruence formed by
oriented twistor null lines and covering the Kerr-Schild spacetime twice.}
\end{figure}

Twist of the congruence determines the complicate form of the Kerr
solution, in spite of the extremely simple form of the metric
(\ref{KS}). One sees also the very important fact that the
Kerr-Schild space-time is twosheeted: the outgoing rays of the
Kerr congruence with tangent vector $k^\m_+$ and the ingoing ones
$k^\m_- $ are positioned on the different sheets of the space and
do not interact with each other.  The Kerr-Schild formalism
\cite{DKS} takes into account that electromagnetic solutions have
to be aligned only to one of two congruences, i.e. the time
orientations of the electromagnetic and gravitational fields have
to be matched. It meets the requirements of holographic principle
\cite{Hoof}, since space is projected by the twistor light-like
rays on the disk $r=0$ (``Radial'' coordinate of the Kerr-Schild
oblate spheroidal system) which separates the positive and
negative sheets of the Kerr space, and also corresponds to the
presumable properties of a quantum black hole space-time
\cite{StHW}, allowing to separate the in- and out- vacua and to
consider the process of evaporation as a  scattering of the
in-vacuum on black hole.

Elasticity of the horizon \cite{BurA} follows from the form of
function $H ,$  \cite{DKS},  \be H =\frac {mr - |\psi|^2/2} {r^2+ a^2
\cos^2\theta} , \label{Hpsi} \ee where the function $\psi(Y)$
determines electromagnetic field. In the Kerr-Schild class of the exact stationary solutions, function
$\phi$ may be {\it any holomorphic} function of the complex
angular coordinate $Y.$ \fn{Function $Y(x)$ determines also the Kerr-Schild tetrad $e^a$:
$e^1 =d\zeta - Y dv, \quad e^2 = d\bar\zeta -\bar Y dv, \quad e^3 = P
k_\m dx^\m, \quad e^4 =dv + h e^3, \quad h=HP^{-2},$ where  $
 P=(1+Y\Y)/\sqrt{2}.$}

\section{Exact Kerr-Schild solutions with singular beams}
The famous Kerr-Newman solution is the simplest solution of the
Kerr-Schild class having $\psi=q=const.,$ where $q$ is the value
of charge. However, any holomorphic function $\psi(Y) $ yields
also an exact solution of this class \cite{DKS}. It is known that
 any holomorphic function on sphere, if it is not a constant, should have
at least one singular point. Therefore, {\it all the exact Kerr-Schild
solutions, except the Kerr-Newman one, acquire one or more
lightlike singular beams} which are positioned along the lines of
Kerr congruence, while the usual harmonic solutions with smooth
angular dependence are absent on the Kerr background.  Vector
potential of electromagnetic field has the general form
\be \alpha =\alpha _\m dx^\m \\
= -\frac 12 Re \ [(\frac \psi {r+ia \cos \theta}) e^3 + \chi d \Y
], \label{alpha} \ee where $ \chi = \int P^{-2} \psi dY  $ and
$\bar Y$ being kept constant in this integration.
 The expression $dY$ represents gradient of the
complex surfaces $Y=const.$  obeying the conditions \be
Y,_2=Y,_4=0 . \label{Y24} \ee These surfaces are totally null,
spanned by the tetrad forms $e^1$ and $e^3.$ Similarly, $d\Y$ is
spanned by $e^2$ and $e^3 ,$ and therefore, $\alpha^\m$ is spanned
by $e^1, \ e^2$ and $e^3 .$ So, using the null tetrad
orthogonality relations $(e^1)^2=(e^2)^2=(e^3)^2=0$ and $e^1
e^3=e^2 e^3=0$ and $e^3 ,$ one obtains that vector potential
satisfies the alignment condition $\alpha _\m e^{3\m} = P \alpha
_\m k^\m=0 .$ The `elementary' beams, formed by a single pole
$\psi(Y) = q / (Y-\hat Y) $ at the point $\hat Y \in S^2,$
propagate along the line of Kerr congruence in direction $k^\m$
corresponding to $Y=\hat Y.$ They play exceptional role, turning
in the far zone (see \cite{BurAxi}) into uniform string-like
singular pp-wave (A.Peres)  solutions, \cite{KraSte}, which have
very important quantum properties, being exact solutions in string
theory with vanishing all quantum corrections \cite{HorSte}.

In general, holomorphic function may be expanded in the Loran series
containing a singular part
$\psi(Y) = \sum_{n=-\infty }^0 q_n Y^n ,$ which can be represented by a series of the above
poles, and a regular one
$\psi(Y) = \sum_{n=0}^{\infty } q_n Y^n .$
We shall see later that the regular polynomial part plays very important role in the
nonstationary solutions.
 It was shown, that singular light-like beams deform topologically
  the horizon, forming the holes connecting the internal and external regions of black hole,
  see \cite{BEHM2}.
The `elementary' single-pole solution may trivially be extended to the case of arbitrary
number of single poles
 \be\psi (Y) = \sum _i \frac {q_i} {Y-Y_i}, \label{Yi}\ee
 in different directions $Y_i =e^{i\phi_i}\tan \frac {\theta_i} 2 .$
Elementary excitation $\psi_i(Y)= q_i /(Y-Y_i) ,$ describes a singular light-like beam
(pp-string) along the null ray of the Kerr congruence in direction
$k_i^\m=k^\m (Y_i, \Y _i).$ The vector potential (\ref{alpha}) is trivially
generalized to a sum over
beams, where for the single i-th beam $
\chi_i= q_i P_i^{-2} \ln(Y-Y_i) + const. + {\cal O} (Y-Y_i) ,$ and
$P_i=(1+Y_i \Y_i)/\sqrt{2} .$
 The corresponding vector field
(\ref{alpha}) gives rise to electromagnetic field $f = \frac 12
F_{ab} e^a\wedge e^b =-d\alpha $ which is aligned with the Kerr
congruence, \be \alpha_\m k^\m =0, \quad  k^\m F_\m ^\n = \lambda
k^\n , \label{align}\ee where $ \lambda = Re \ [\psi
/(r+ia\cos\theta)^2 ].$
It was shown that these beams have strong back reaction on the metric, via
function $\psi_i(Y)$ entering in (\ref{Hpsi}).
The multibeam solution (\ref{Yi}) is a particular case of the exact solutions
obtained by Debney, Kerr and Schild (DKS) in seminal work \cite{DKS}.
It should be emphasized that in the usual perturbative approach the beam solutions
are absent, because the important alignment condition are dropped out of the
perturbative equations, and a mixing of the in and out fields occurs on the same
sheet of metric.

The appearance of light-like beam pulses is a pure classical effect,
however, it allowed us in \cite{BEHM2} to put some conjectures concerning semiclassical
treatment of black hole of interaction with weak electromagnetic field and, in particular,
with electromagnetic vacuum. Since the black hole horizon is extra sensitive  to electromagnetic
excitations of black hole, it  may also be sensitive to zero point field (ZPF) which
is exhibited classically as Casimir effect. Therefore, as it has been discussed by many
authors, vacuum fluctuations have to generate fluctuations of the metric and horizon.
Note that such point of view assumes tacitly that there exist some semiclassical
pre-quantum geometry which lies beyond the usual classical gravity and describes the
back reaction of the vacuum fluctuations to metric.

\begin{figure}[ht]
\centerline{\epsfig{figure=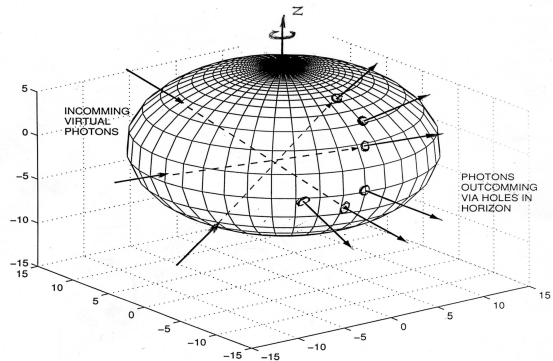,height=6cm,width=8cm}}
\vspace*{-7mm}
 \caption{Excitation of a black hole by the
zero-point field of virtual photons forming a set of micro-holes
at its horizon.}
\end{figure}

The exact Kerr-Schild solutions show that the usual smooth harmonic basic
functions are absent on the Kerr background, giving the way to
singular beams forming an overfilled system of the coherent
non-normalizable states. It rises the questions on the finiteness
of the energy of the Kerr-Schild beam-like  solutions. Considering the
singular beams as a fine-grained vacuum structure, we should take
into account that the electromagnetic vacuum energy is divergent, and a
transfer to the classical Einstein-Maxwell theory demands at least
a regularization of the stress-energy tensor. For example, for
calculation of the Casimir energy, de Witt uses the regularization
by subtraction \cite{deWit}, $ T^{(reg)}_{\mn} = T_{\mn} -
<0|T_{\mn}|0>,$ under the condition $ T^{(reg) \ \mn} ,_\m = 0 .$
As opposed to a cumulative action of the vacuum in Casimir effect,
the treatment of the fine-grained structure of fluctuations may
require another form of the regularization. As we shall see, the
structure of the Debney-Kerr-Schild equations (DKS), \cite{DKS},
suggests a regularization which retains the fluctuating fine-grained
structure of the horizon and the semiclassical pre-geometry.
To transfer to the usual classical gravity with a smooth background
geometry and get the usual scenarios of collapse, an extra regularization
of the fine-grained structure is necessary.

  The real beams has to be finished at some distant black holes or particles,
or at the infinitely distant matter. It
 is confirmed by the treatment of multicenter Kerr-Schild solutions,
 \cite{Multiks}, in which the similar beams appear between the
 distant sources, being supported by a twistor line common for them.

The exact pp-wave Peres solutions may have a carrier frequency and
a finite extension. Therefore, it is desirable to consider a
minimal generalization of the exact stationary Kerr-Schild beam-like
solution to time-dependent beam pulses, which is necessary to
consider a time-dependent process of scattering.

\section{Time-dependent Kerr-Schild solutions} The closest time-dependent
generalization to (\ref{Yi}) is given by the form \be \psi(Y, \t)
=\sum_i c_i(\t) (Y-Y_i)^{-1}, \label{psinst}\ee where one
assumes that an elementary beam has $c_i(\t)=q_i(\t)e^{i\omega_i
\t }$ where  $q_i(\t)$ is amplitude and $\omega_i$ is carrier
frequency.
 Existence of the exact
nonstationary solutions of this type is a priory problematic, and
the problem may has two stages: i) exact solutions for electromagnetic field on
the Kerr background, ii) self-consistent solutions, taking into
account the back reaction of the electromagnetic field on metric. General
equations for nonstationary electromagnetic Kerr-Schild solutions were obtained by by
Debney, Kerr and Schild (DKS) \cite{DKS} in 1968, and the general
electromagnetic solution on the Kerr-Schild background was obtained only in 2004,
\cite{BurAxi}.
 The analyzed in \cite{BurAxi} wave solutions contained singular beams along the
$\pm z-$ half-axis, and there appeared the conjecture that, for
seldom exclusions, the exact nonstationary electromagnetic
solutions on the Kerr background should acquire the beams. Similar
to stationary case, the nonstationary Kerr-Schild solutions
contain the singular beams pulses which change the structure of
black hole horizon. The considered bellow important peculiarities
of the nonstationary solutions are determined  by a specific
structure of the DKS equations. electromagnetic field is described
by the self-dual tetrad components, \be\cF _{12}=AZ^2, \quad \cF
_{31}=\gamma Z - (AZ),_1, \label{E6}\ee where $\cF_{ab}= e_a^\m
e_b^\n \cF_\mn ,$ and the function $Z$ is a complex expansion of
the congruence, $Z=Y,_1 .$ For the Kerr-Newman solution at rest
$Z$ is inversely proportional to a complex radial distance  $ Z=
-P /(r+ia\cos \theta).$ Here $P$ is a conformal factor which is
determined by Killing vector of the solution \cite{DKS}. For a \bh
at rest $P=2^{-1/2}(1+Y\Y).$ \fn{We neglect the recoil, $\dot P=0
,$ assuming that that the mass of  \bh is much greater then the
energy of excitation.} The equations for function $A$ are the same
as for stationary case $ (AP^2),_2 = 0,\ A,_4=0 .$ They have the
general solution $ A= \psi P^{-2} \label{A} ,$ where function
$\psi$ has to satisfy \be \psi,_2=\psi,_4=0 . \label{psi24}\ee So,
in the nonstationary case, there appears the unique difference
that function $\psi$ has to depend on extra retarded-time
parameter $\t ,$ which must obey the equations $ \t,_2=\t,_4=0 . $
The principal difference from the stationary case is contained in
the second electromagnetic DKS equation which in \cite{BurAxi} was
reduced to very simple form \be  (\gamma P),_\Y = - \dot A
\label{EMgamma} ,\ee showing that {\it any nonstationarity in
electromagnetic field ($\dot A\ne 0$) generates an extra function
$\gamma $} which, in accord with (\ref{E6}), generates also the
lightlike electromagnetic radiation along the Kerr congruence.
Such a radiation is well-known for the Vaidya `shining star'
solution \cite{KraSte}, in which the field $A=\psi P^{-2}$ is
absent and $\gamma$ is incoherent, being related with the loss of
total mass into radiation, \be \dot m = - \frac 12 P^2 \gamma \bar
\gamma . \label{G2}\ee

This is one of two gravitational
equations determining self-consistency of the Kerr-Schild solution \cite{DKS}.
The most important consequence following from DKS equations in the nonstationary
case is the fact that the field $\gamma$ appears inevitable, however it does
not contribute to deformation of the horizon, since it is absent in the function
$H$ of (\ref{Hpsi}). Its back reaction on metric is smooth and circumstantial,
acting only via the slowly decreasing mass parameter $m .$
Therefore, in the nonstationary Kerr-Schild case we obtain that the
{\it lightlike fields, determined by functions $\psi$ and $\gamma ,$ have
essentially different impact on the horizon.}

Function  $\psi= \psi(\t, Y)$ obeys the equation (\ref{psi24})
which shows that the retarded time $\t$ has to satisfy the
conditions similar to (\ref{Y24}), and therefore, gradient of $\t$
has to be aligned to congruence, \be k^\m \t,_\m =0 . \ee It was
obtained in \cite{BurAxi} that the corresponding retarded-time
parameter has the form

\be \t = t -r -ia \cos \theta ,\ee  and the
general solution of the equations (\ref{EMgamma}) takes the form
\be
\gamma = \frac{2^{1/2}\dot \psi} {P^2 Y}
  +  \phi (Y,\t)/P ,
\label{12}\ee where $\phi (Y,\t)$ is the second arbitrary analytic
function of $Y$ and $\t $ which is solution of
homogenous equation (\ref{EMgamma}) corresponding to $\dot A =\dot \psi=0 .$

Since function $\phi$ contributes only to $\gamma$ it does not
impact on the form of the horizon too. Important role of this
function is obtained from analysis of the second gravitational Kerr-Schild
equation [eq.(5.44) in \cite{DKS}] which is reduced to the form
(see \cite{EM3p}) \be m,_\Y = \psi\bar \gamma P . \label{G1}\ee

If we note that
$\gamma \sim \dot \psi \approx i\omega \psi ,$ we obtain that the
r.h.s. of this equation tends to zero in the low-frequency limit,
as well as the r.h.s of the equation for $\dot m .$ So, the full
solutions tend to exact ones (consistent with gravity) in the
low-frequency limit \cite{EM3p,BEHM3}.

However, there is a more consequent way for interpretation of these solutions
which is close to a quantum version of the Einstein-Maxwell equations.
So far as we consider  the vacuum electromagnetic field, one has to assume that electromagnetic
field should have an operator meaning and a regularization of the stress-energy
(r.h.s. of the gravitational DKS equations) is necessary.

Note that the Vaidya `shining star' solution \cite{KraSte} corresponds to
$\psi =0,$ and the equation (\ref{G1}) shows that $m$ is independent from $Y,$
 while the field $\gamma$ in (\ref{G2}) is
assumed to be incoherent and has to be considered with averaged r.h.s.,
 $\frac 12 <P^2 \gamma\bar \gamma> .$  This
approach can be extended to the r.h.s of the both gravitational
equations which acquire the form \be m,_\Y = <P\psi\bar \gamma> ,
\quad \dot m = - \frac 12 <P^2 \gamma \bar \gamma>, \label{averG}\ee
and may be considered as a semiclassical analog of  quantum
approach. \fn{In \cite{KobTom} the general Kerr-Schild
solution (\ref{12}) is applied to a black hole-disk system and some
interesting properties of the averaged stress-energy tensor are
observed.}

 Since function $\phi$ is free, its form and parameters (positions of poles)
of the function $\phi/P$ may be tuned to cancel the poles of function
 $\dot \psi = \sum _i \dot c_i(\t)/(Y-Y_i) $
  in (\ref{12}). We set
\be \phi^{(tun)}_i (Y,\t) = - \frac{2^{1/2}  \dot c_i(\t)}
{Y (Y-Y_i) P_i }, \label{itunphi} \ee where
\be P_i = P(Y,\Y_i) =
2^{-1/2}(1+Y\Y_i)\label{Pi} \ee is analytic in $Y ,$ which
provides required analyticity of $ \phi^{(tun)}_i (Y,\t) .$
Using the equality
\be (P_i - P)/(Y_i-Y) =\frac {Y} {\sqrt{2}}  \frac {(\Y_i-\Y)} {(Y_i-Y)} ,\ee
we obtain that the regularized field
\be \gamma_{(reg)} = \frac{2^{1/2}\dot \psi} {P^2 Y}
  +  \sum_i\phi^{(tun)}_i (Y,\t)/P
\label{gamreg}\ee

 takes the form
\be \gamma_{(reg)} = \frac 1{P^2}\sum_i \frac {\dot c_i} {P_i}
[\frac {\Y_i-\Y}{Y_i-Y}] .\ee It should be mentioned that
$\gamma_{(reg)}$ is exact regular solution of (\ref{EMgamma}).
Therefore,  the exact solutions of DKS equation (\ref{EMgamma})
contains remarkable subclass of self-regularized solutions
$\gamma_{(reg)}$.

R.h.s. of the equation (\ref{G1}) takes the form

\be \nonumber \psi\bar \gamma_{(reg)} P = \sum_{i,k} \psi_i
\bar\gamma_{k (reg)} P =
 \sum_{i,k}   \frac{ c_i\dot {\bar c}_k (Y_k -Y)} { P \bar
P_k (Y-Y_i)(\Y-\Y_k)} , \ee

where the terms with $i\ne k$ are not correlated and drop out
after averaging. The rest is

\be < \psi\bar \gamma_{(reg)} P > =
 -\sum_{k}   \frac{ c_k\dot {\bar c}_k } { P \bar
P_k (\Y-\Y_k)} . \ee

Equation (\ref{G1}) may be integrated
using the Cauchy integral formula, and we obtain
the expression
\be < m >_t = m_0 - 2\pi i  \sum_k \frac{c_k\dot {\bar c}_k }
{|P_k|^2} \label{taver}\ee
containing contributions from the residues at singular points $Y_.$

When $c_i(\t)$ is expressed
 via amplitudes $q_i(\t)$ and carrier frequencies $\omega_i ,$
$c_i(\t)=q_i(\t)e^{-i\omega_i \t } ,$ the impact of the carrier
frequencies disappears, \be <m>_t = m_0 + 2\pi \sum_k \omega_k
\sum_{k} <\frac{ q_k {\bar q}_k } { | P_{kk}|^2 }> ,
\label{averm}\ee however, the mass term retains slow fluctuations and
an angular non-homogeneity  caused by casual distribution
of the beams in different angular directions.

It is known that the second gravitational equation in
(\ref{averG}) is really a definition of the loss of mass in
radiation. The time averaging removes again the terms with $i \ne
k$  and yields

\be <\dot m>_t  =  - \frac 12 \sum_{k}\frac{\dot c_k\dot {\bar
c}_k}{P^2  |P_{k}|^2} \label{averrad}. \ee

In terms of the amplitudes of beams we obtain
 \be < \dot m >_t = - \frac 12 \sum_k \omega_k^2 <\frac{ \dot q_k {\bar q}_k }
{|P_{kk}|^4} > ,\ee which shows influence of the single beams to
the mass evaporation. These sporadic fluctuations of the mass term
caused by the individual beam pulses may also be averaged over the
time and angular directions, however, this smoothing does not
remove the sharp back-reaction of the beams to metric and horizon
produced by poles in $\psi(Y,\t)$ in agreement with (\ref{Hpsi}).

Therefore,  the obtained solutions are exact for the
time-dependent electromagnetic field on the Kerr-Schild background
and, up to our approximation which neglects the recoil, the
obtained solutions are consistent with the Einstein-Maxwell system
of equations with averaged stress-energy tensor.

\section{Holographic gravity of the time-dependent KS solutions}

 The obtained very broad class of semi-classical solutions to Einstein-Maxwell
 equations have two principal peculiarities:
 the solutions are time-dependent and contain a sporadic flow of beam
 pulses. Therefore, the  solutions reveals a
 classical fine-grained fluctuating geometry of twistor-beams and
 exhibit a mechanism of evaporation provided by the specific
 structure of DKS equations. They show explicitly that
the outgoing radiation contains two components determined by
functions $\psi$ and $\gamma=\gamma_{(reg)} .$ Both the functions
are creating the null electromagnetic radiation along the twistor
null lines of the Kerr congruence in agreement with (\ref{E6}),
however, they play their own specific role:

a) function $\psi $ describes a casual set of the outgoing
singular beams which perforate the horizon, forming fluctuating
micro-holes  breaking impenetrability of the horizon.

b) the field determined by function $\gamma_{(reg)}$ is regular
and akin to the Vaidya `shining star' radiation. In the agreement
with $\dot m = - \frac 12 P^2 \gamma_{(reg)} \bar \gamma_{(reg)}
,$ it is responsible for the mass evaporation.

Therefore, the resulting evaporation represents a classical analog
of the quantum tunnelling process.

Holographic interpretation of the mysterious twosheetedness of the
Kerr-Schild geometry allows one to treat evaporation as a scattering
and reveals the important role of the  `negative' sheet of the Kerr
geometry, showing that Kerr's twosheetedness represents a
classical analog of  the required holographic structure of a quantum
black-hole space-time predicted by Stephens, t' Hooft and Whiting
(StHW) in \cite{StHW}. The holographic StHW space-time is to be
divided into two causally-related `in' and `out' regions joined by
a 2+1 dimensional (shell-like) boundary which is holographically
dual to the `in' and `out' bulk regions. Twosheeted structure of
the Kerr geometry is perfectly adapted to holographic StHW structure.
It has to be unfolded, forming the in and out bulk regions separated by the
Kerr BH source, as it is shown on the Penrose conformal diagram,
Fig.3.

\begin{figure}[ht]
\centerline{\epsfig{figure=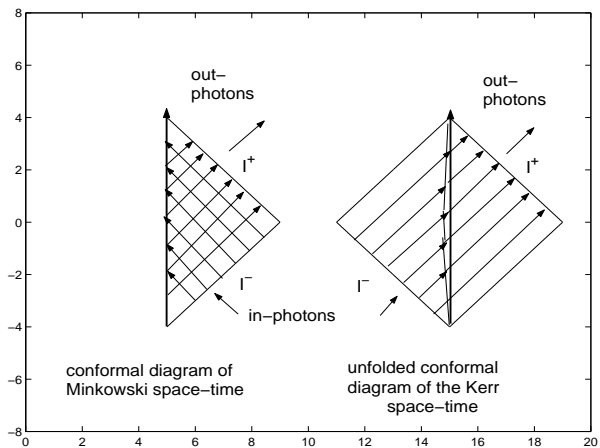,height=6cm,width=8cm}}
\caption{Penrose conformal diagrams. Unfolding of  the auxiliary
$M^4$ space of the Kerr spacetime to
 StHW structure of a quantum BH spacetime. }
\end{figure}

The BH appears as a holographic image created by the projection
performed by twistor rays from past infinity $I^- .$ After
scattering on the Kerr BH, the out-going beams propagate along the
twistor rays of Kerr congruence to future infinity $I^+ .$
Therefore, radiation from BH appears as a result of scattering of
the in-vacuum field on the shell-like source of the Kerr BH which
may be, in particular, a shell of collapsed matter positioned very
close to horizon, or a matter already collapsed under the horizon
of an even BH, either also a disk-like source of the Kerr-Newman
geometry without horizon \cite{Isr,Lop,RenGra,BurBag,BEHM}, as it
was considered in the models of the Kerr spinning particle.
Similar, authors of \cite{StHW} argue that such a source may also
be positioned inside of the horizon and the existence of the
horizon, in principle, is not important for derivation of the BH
radiation. In all the cases the Kerr source may be replaced by a
rotating shell which separates the `in' and `out' bulk regions of
the Kerr space-time. In the simplest case it is the disk $r=0$
spanned by the Kerr singular ring - the Israel-Hamity Kerr's
source \cite{Isr}.
  Since the twistor rays of the Kerr congruence are
time oriented null lines, the Kerr-Schild alignment condition
(\ref{align}) plays specific role of time ordering on the
considered semi-classical level. However, the Kerr-Schild
solutions are analytic solutions which may be extended
analytically back onto negative sheet of the Kerr space, and the
treatment of the StHW Kerr-Schild structure together with the
holographic conception of scattering, assume that this analyticity
is broken by the shell-like Kerr source. It allows us to propose
that the in-sheet, $r<0 ,$  is to be residence of the vacuum
in-fields
 $\gamma_{in} .$ In some sense this sheet may be considered  as
 nonphysical one, since the black hole horizon is absent  by $r<0 .$
 One can also suppose that the in-field $A_{in}$ is also absent
 there (at least, we has to put $\dot A_{in}=0$), which yields
 $\gamma_{in}= - \gamma_{(tun)} .$ Thus, we obtain that the `negative'
 in-sheet is a region of the homogenous solutions of DKS equations
 which may contain singular beams but do not produce singular back
 reaction to the metric. Therefore, the metric turns out to be
 smooth by $r<0 .$
 The scattering of the field $\gamma_{in}$ at the Kerr source creates
 the outgoing electromagnetic singular beams  $A_{out},$ accompanied
  by singular deformations of the metric and horizons and by the
  regular outgoing incoherent thermal field
  $\gamma_{out}=\gamma _{reg} .$

The holographic bounce conception \cite{StHW}, related with the
treatment of the BH evaporation as a scattering,  has the specific
feature that the beams are scattered on the 2+1 dimensional
shell-like source. Taking into account, that the solutions on the
in- and out- Kerr's sheets have
 in this case their own doubling by analytic extensions to another half-sheet,
 we arrive at a possible interplay of the four half-sheets. Moreover, from
the treatment of multi-center Kerr-Schild solutions
\cite{Multiks}, we know that any extra matter source adds two more
sheets to the KS geometry. Therefore, the presence of the matter
shell source adds at least two more sheets to the KS space-time
and have to lead to the KS solutions with a shell-like
singularity. Therefore, the usual Greene functions have to be
generalized to some KS solutions which are singular at the shell.
It is clear that topological interplay of these half-sheets may be
related with the origin of nontrivial commutation relations. These
questions need especial treatment which goes out of the frame of
this work.

The next important  problem is related with spectrum of the
evaporation. The obtained analytical solutions have continuous
spectrum, displaying that the discrete quantum spectrum can appear
only as a result of interaction of the in-vacua with a matter
source of the BH.

The simplest source of the even Kerr BH is the disk positioned at
$r=0 .$  In fact, it has a stringy structure, representing a
light-like closed Kerr string positioned on the boundary of the
Kerr disk \cite{Isr,Bur0,BurBag,BurOri,BurTwi}. This string is
reminiscent of the fundamental heterotic string \cite{BurSen}, the
Sen solution to low energy string theory  \cite{Sen}.

The averaged stress-energy tensor of the Kerr-Schild solutions,
\be T_{(\gamma)}^{\m\n} = Z\bar Z <\gamma_{(reg)}\bar
\gamma_{(reg)} P^2> k^\m k^\n  ,\ee is determined by term
$\gamma_{out}=\gamma_{(reg)}$ and is independent from the position
and even from the presence of the horizon. It has axial symmetry
and the symmetry of time-translations which characterize the
symmetry of a conformal field theory determined by the function
$\gamma_{out} (Y, \t).$ For the Kerr geometry at rest $Z=P/(r+i a
\cos \theta).$ Considering the disk $r=0$ as a holographically
dual boundary, we obtain that the stress-energy tensor turns out
to be singular in the equatorial plane $\theta=\pi /2 ,$ which
corresponds to the CFT on the Kerr singular ring, $r=\cos \theta=0
,$ i.e. CFT of the Kerr light-like string \cite{BurOri,BurSen}.
Therefore, the quantum modes of the excitations (in this case only
one side movers) appear  as a result of resonance of the in-vacuum
field on the singular string-like source of the Kerr BH. It gives
evidences that the obtained radiation of Vaidya type corresponds
to Hawking radiation in the agreement with two-dimensional CFT of
the string theory related with the Kerr singular ring. One can
chose $\d_\phi$ and $\d_t$ as generators of a Virasoro algebra for
the spacetime diffeomorphisms associated with CFT on the cylinder
 $(\phi, t) ,$  and following \cite{Carl,Stro} use the Cardy
 formula for spectrum. \cite{Carl}.
 Formally, it is similar to the given by Strominger derivations for the
 boundary CFT positioned at infinity, but we have absolutely
 different physical mechanism related with excitations of the Kerr string.

 In the many recent works, for example \cite{Solod,Star,KerCFT}, the
 holographically dual shell source represents a dense matter
 concentrated near the horizon of the Kerr BH. All the above features
 of the bounce StHW model are retained for exclusion of the clear
 exhibition of the stringy structure of the Kerr source. The
 horizon of rotating BH takes the form of a rigidly rotating disk,
 oblateness of which depends on the velocity of angular rotation,
 and a string-like region is formed on the edge rim of the disk
 only for the quickly rotating extremal black-holes. It all
 other cases the above sharp stringy effect turns out to be
 smeared, and presumably, spectrum may be formed as a mixing
 of spectrum for many strings.

Details of the quantum version of the  holographic Kerr-Schild
space and the thermal processes related with the quantum resonance
demand extra analysis and have to be considered elsewhere.

It should also be noted  that in the obtained solutions we neglected
by the recoil, presence of which essentially complicates DKS
equations and leads also to extra problems with geometrical
matching of the in and out sheets. There appears also a lot of other
important and interesting questions which we have to leave
for further considerations.

\section{Conclusion}

 The obtained Kerr-Schild solutions and the holographic representation
 of the Kerr twosheetedness showed that the Kerr-Schild geometry is
 a semiclassical analog of the holographic StHW  space-time suggested
 for quantum BH background.
 The obtained solutions are time-dependent and describe a fine-grained
 fluctuating holographic BH geometry  which takes an intermediate
 position between the classical and quantum gravity. Basic elements of this
 geometry are beam pulses which take asymptotically the form of pp-waves.
 Singular lines of the beams are supported by twistor null lines of the
 Kerr-Schild geometry. This fine-grained structure of spacetime formed by twistor
 beams represents a holographic alternative to the considered in \cite{Hoof} lattice
 structure of vacuum formed by giant masses positioned at the nodes of the
 lattice.

 Therefore, along with the old Penrose statements on the principal role of twistors
 in the structure of space-time, there appear extra evidences that
 the oriented twistors may be considered as basic elements of the vacuum
 structure. Principal advantage of the twistor vacuum structure with respect to the
 usual lattice structure is that the vacuum based on twistors possesses the
 explicit Lorentz invariance and  explicit time orientation.
Twistor description of the massless fields based on the WZW or
topological B-model represents the basis of the Nair-Witten
concept on the scattering of the gauge amplitudes in twistor space
\cite{Nai,Wit}. The Penrose claim that twistors may be considered
as primary objects of 4D space-time with respect to the space-time points
(which may be considered as geometrically dual to
twistors) is explicitly illustrated by the holographic
structure of the Kerr-Schild black hole space-time which allows one to
describe the structure of the Kerr black-hole via the Kerr congruence of
twistor null lines.

\section{Acknowledgement}

Author would like to thank K.S. Demirchan, E. Elizalde, S.R. Hildebrandt , G.
Magli,  T. Nieuwenhuizen and D. Singleton for useful discussions.

\end{document}